\documentclass[multphys,vecphys]{svmult}

\usepackage{makeidx}         
\usepackage{graphicx}        
\usepackage{multicol}        
\usepackage[bottom]{footmisc}

\makeindex             

\newcommand {\hi} {{\rm H\,{\small I}}~}

\begin{document}

\title*{The bright and the dark side of Malin 1}

\author{Renzo Sancisi\inst{1,2}
\and
Filippo Fraternali\inst{3}}


\institute{INAF, Observatory of Bologna, I
  \texttt{renzo.sancisi@oabo.inaf.it}
  \and 
  Kapteyn Astronomical Institute, Groningen, NL
  \and
  Department of Astronomy, University of Bologna, I
  \texttt{filippo.fraternali@unibo.it}}

\maketitle

\begin{abstract}

Malin 1 has long been considered a prototype giant, dark matter dominated Low
Surface Brightness galaxy.
Two recent studies, one based on a re-analysis of VLA \hi observations and 
the other on an archival Hubble I-band image, throw a new light
on this enigmatic galaxy and on its dark/luminous matter properties.

\end{abstract}

\section{Introduction}
\label{sec:1}

Malin 1 is a highly unusual disk galaxy characterized by an enormous, 
\hi rich and extremely low surface brightness disk \cite{bot87, pic97}.
Recent, deep R-band data by Moore \& Parker \cite{moo07} show an exponential 
disk extending out to 124 kpc $h_{[75]}^{-1}$ (scale length 53 kpc). 
This corresponds to the \hi extent.
There is also a prominent {\it bulge}-like component.
According to Pickering et al. \cite{pic97} the rotation curve has the slowly 
rising shape typical of the less luminous, ``dark matter dominated'' LSB 
galaxies.
This slow rise of the rotation curve in the presence of a luminous central
component is in marked contrast with the 
rule that there is a close correlation between the distribution of light and 
the shape of the rotation curve \cite{san04}. 
However, Pickering et al. \cite{pic97} do point out that their rotation 
curve is very uncertain because of the low resolution of the observations, 
the low signal/noise ratio and the strong warping.
This has induced us to carry out a re-analysis of the \hi data with special 
attention for the beam-smearing effects and to make a new comparison with the 
luminosity profile.
The results of our analysis agree with those of a recent 
HST optical study of the bright central component \cite{bar07}:
Malin 1 is a normal, early-type galaxy 
surrounded by a huge, low-surface-brightness outer disk.

\begin{figure}
\centering
\includegraphics[height=9cm]{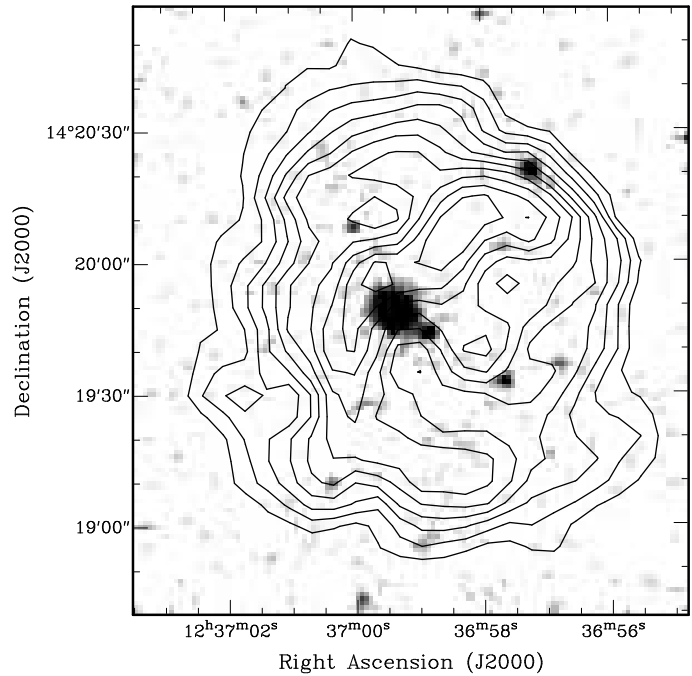}
\caption{Total \hi map of Malin 1 overlaid on the optical DSS image.
Contours run from 7.7 to 49.1$\times 10^{19}$ atoms cm$^{-2}$ with an
increment of 4.6$\times 10^{19}$ atoms cm$^{-2}$. The HPBW is 21'' ($\sim$29 
kpc).}
\label{fig:1}       
\end{figure}

\section{New rotation curve and comparison with the luminosity profile}
\label{sec:2}

The \hi data cube obtained with the VLA by Pickering et al. \cite{pic97} 
has been re-analyzed.
The large extent of the \hi disk is shown in Fig.\ 1 superposed on the optical 
(DSS) image. 
The \hi radius corresponds approximately to that of the extended, faint 
optical disk. 
A new velocity field has been derived. The 21-cm line profiles are strongly 
affected  by beam smearing and are very asymmetric. 
Instead of the intensity-weighted mean velocities used by Pickering et al., 
which suffer heavily from 
beam smearing, we have taken the velocities at the profile peaks, close to 
the high rotation velocity side. 
Subsequently, the rotation curve has been derived from the velocity field 
following well-known standard procedures. 
This rotation curve has been used 
to construct model data cubes to verify its correctness
(Fraternali \& Sancisi, in preparation).
The new and the old \cite{pic97} rotation curves are shown in Fig.\ 2 (bottom).
Amplitude and flat outer part are the same. In the inner parts the new curve 
rises much more steeply and reaches higher values inside
20 arcsec ($\sim$30 kpc) in correspondence of the central concentration in the 
luminosity profile (Fig.\ 2, top).
Fig.\ 3 shows the ``maximum disk'' decomposition, with isothermal 
halo and \hi disk. 
The R-band profile \cite{moo07} has been used. 
The maximum disk M/L ratio is 5.2. 
This is in the range of the values found 
for luminous early-type galaxies \cite{noo07}.

\begin{figure}
\centering
\includegraphics[height=11cm]{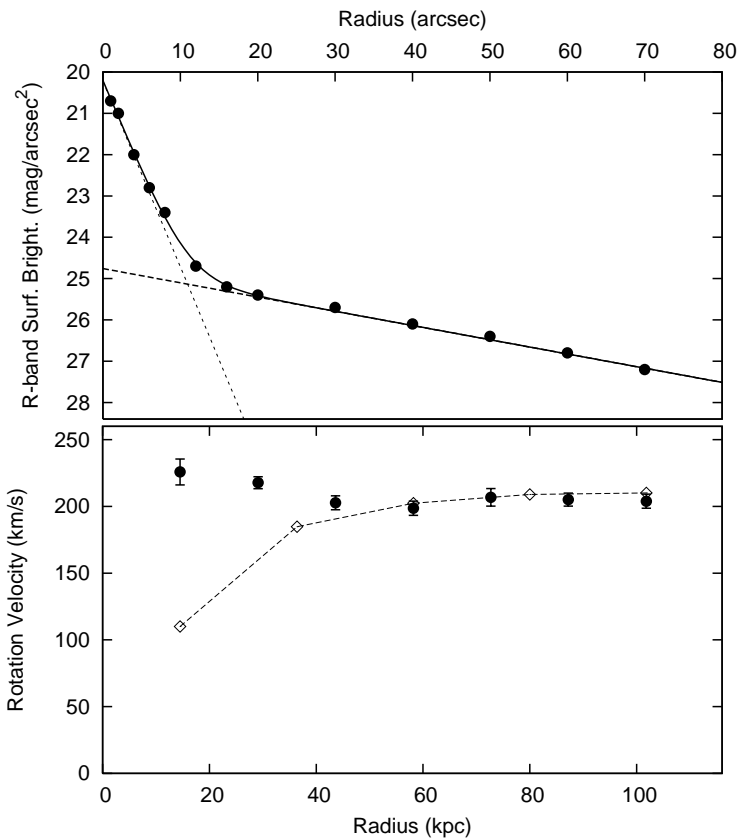}
\caption{Mass follows light.
Upper panel: R-band luminosity profile derived from Moore \& Parker \cite{moo07}
and fitted with two exponential disks.
Lower panel: new rotation curve of Malin 1 derived as described in the text
(filled circles).
The open diamonds show the rotation curve derived by Pickering et al. \cite{pic97}.}
\label{fig:2}       
\end{figure}

\begin{figure}
\centering
\includegraphics[height=6.5cm]{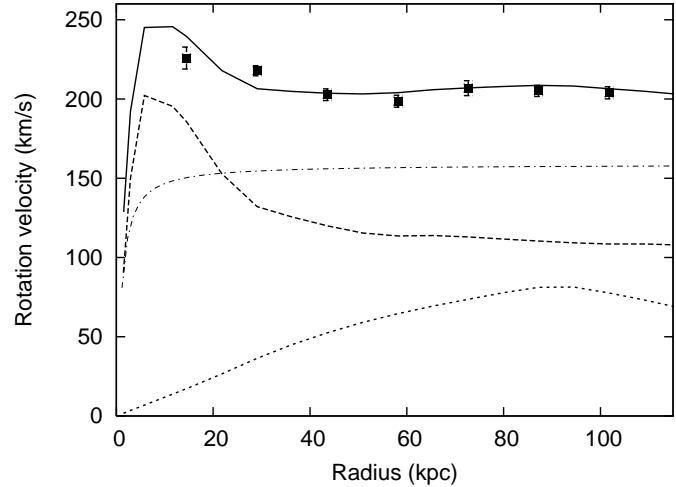}
\caption{New rotation curve (squares) of Malin 1 and ``maximum disk'' 
mass decomposition. 
The contributions from the stellar (thick dash) and gaseous (short dash) 
disks and the DM isothermal halo (dot-dash) are shown.
The thick line shows the sum of the three.}
\label{fig:3}       
\end{figure}

\section{Conclusions}
The present study is based on a re-analysis of existing \hi observations of 
Malin 1 obtained with the VLA. 
A new rotation curve has been derived.
This rotation curve shows a close correlation with the luminosity distribution,
in line with the ``rule'' suggested by Sancisi \cite{san04}.
Also in this galaxy the mass seems to follow the light.
The rotation curve has the shape (steep inner rise) typical of 
high surface brightness (HSB) galaxies. 
The classical disk-halo decomposition of 
the rotation curve has shown that a maximum disk solution is possible. 
Clearly, in its inner luminous part, Malin 1 has the characteristics of 
an early-type HSB galaxy. 

Barth \cite{bar07} has recently published a study of Malin 1 based on 
archival Hubble I-band data. 
He has examined the structure and the properties of the 
inner bright parts and has concluded that Malin 1 has a normal stellar disk 
and that, out to a radius of $\sim$10 kpc, its structure is that of a typical 
SB0/a galaxy.  

The new \hi analysis and the optical study throw
new light on Malin 1 and on its dark/luminous matter properties. 
Both point at the same conclusion: Malin 1 is a normal, luminous early-type
galaxy.
The enigma of the huge (120 kpc), low-surface-brightness stellar and \hi disk 
surrounding the bright inner parts remains. 
In view of the large-scale 
symmetry and regularity and of the large orbital period ($\sim$3.5 billion 
years) in the outer parts, it seems unlikely that the formation of this 
extended structure is due to recent accretion and mergers.

We thank Tim Pickering and his co-authors for kindly making their \hi data cube available to us.

\printindex
\end{document}